\documentclass[5p]{elsarticle}
\usepackage{natbib}
\usepackage{amsmath}
\usepackage{latexsym}
\title{On the dynamics and thermodynamics of terrestrial planets}
\author{Regan L. ~Patton\corref{cor1}}
\address{School of the Environment, Washington State University, Pullman, WA 99164-2812}
\cortext[cor1]{Corresponding author: rpatton@wsu.edu}

\begin{document}

\begin{abstract}
A model for terrestrial planets, inclusive of viscous fluid behavior and featuring finite normal stress differences, is developed. This work offers new insights for the interpretation of planetary survey data.  Evolution equations for poloidal and toroidal motions include gradients of density $\rho$, viscosity $\eta$ and normal stress moduli $\beta_1,\beta_2$.  The poloidal field exhibits gradients in the cubic dilation, which couple non-isotropic pressures to the combined deformation field.  In contrast, the toroidal field exhibits vorticity gradients with magnitudes proportional to the natural time $\smash{\frac{\beta_1}{\eta}}$.  This holds even in the absence of material gradients.  Consequently, viscosity gradients are not required to drive toroidal motions.  The toroidal field is governed by an inhomogeneous diharmonic equation, exhibiting dynamic shear localization. The strain-energy density for this model, as a function of temperature, is found via thermodynamics.  Assuming heat transfer with characteristic diffusivity $\kappa$, a radial model parameterized by thermomechanical competence $\smash{\frac{\kappa}{\chi}}$ is found, where $\chi=\smash{\frac{\eta{l}^2}{\beta_1}}$ is a diffusivity for microphysical dislocations. Shear dislocations, admissible for $\smash{\frac{\kappa}{\chi}>\frac{1}{2}}$, are found to coincide with supershear rupture speeds for in-plane (Mode II) cracks. This range of thermomechanical competence coincides with depths in the crust, upper mantle and transition zone where earthquake foci are observed. Consequently, all seismic sources must exhibit some supershearing component. Observed variations in Earth's gravity-topography admittance and correlation spectra, and earthquake moment-depth release are interpreted in light of this hypothetical structure.        
\end{abstract}


\maketitle
\tableofcontents

\section{Introduction}
In 2006 the International Astronomical Union defined a planet as a celestial body that: (1) orbits around the Sun, (2) is sufficiently massive to achieve hydrostatic equilibrium (i.e. is spheroidally shaped), and (3) has cleared its orbit of other objects.  Pluto, not satisfying the third rule, subsequently has been categorized as a dwarf planet.  The distinction between a dwarf planet and an object of lesser mass, called a small solar system body (SSSB), hinges on the second rule and therefore constrains the minimum mass required for hydrostatic equilibrium.

Most SSSBs in the inner solar system depart from spheroidal symmetry \cite{Lewis2004}.  The two largest, Pallas and Vesta, have mean diameters of $545 km$ and $525 km$, respectively, but only the latter has been surveyed from orbit.  Vesta is spheroidally shaped, except for the Rheasilvia impact basin at its southern pole, and has a mean density of about $3.4 g/cc$ (Table 1).  For comparison, the dwarf planet Ceres has mean diameter $952 km$ and mean density about $2.1 g/cc$.  Assuming sphericity, these data can be used to estimate the masses of these objects as $\smash{M=\frac{\pi}{6}\rho{d}^3}$, where $\rho$ is density and $d$ is diameter.  Consequently, the minimum mass required for hydrostatic equilibrium in typical rocky materials is about $10^{20} kg$, some 3-4 orders of magnitude less than the mass of terrestrial planets.

\vspace{10pt}
Table 1.  Mean densities and orbital semi-major axes for selected planetary mass objects \par
\begin{tabular}{@{}lccl@{}}
Object & $\rho (g/cc)$ & $a (A.U.)$ & Category\\
\hline
Mercury & $5.4$ & $0.39$ & terrestrial planet\\
Venus & $5.2$ & $0.72$ & terrestrial planet\\
Earth & $5.5$ & $1.0$ & terrestrial planet\\
Moon & $3.3$ & $\emph{1.0}$ & rocky satellite\\
Mars & $3.9$ & $1.5$ & terrestrial planet\\
Vesta & $3.4$ & $2.3$ & SSSB\\
Ceres & $2.1$ & $2.8$ & dwarf planet\\
Pallas & $2.8$ & $2.8$ & SSSB\\
\hline
\end{tabular}
\vspace{10pt}

The mean diameter of the largest SSSBs is similar to flexural wavelengths found in regional isostatic studies on terrestrial planets \cite{Watts2001}.  Also, note that Earth's gravity-topography correlation levels off for spherical harmonic degrees $l\geq25$, corresponding to half-wavelengths less than about $800 km$ (Fig.1).  In contrast, the admittance levels off for $l\geq200$, corresponding to half-wavelengths less than about $200 km$.  Curiously, more than $99.9\%$ of SSSBs are smaller than $200 km$ in size.  Is there a common explanation for these trends?

Establishing a clear definition for hydrostatic equilibrium is complicated by rigidity, particularly for rocky or icy objects.  This includes the terrestrial planets which are composed of silicate rocks and metal, and possess solid surfaces and secondary atmospheres.  The strength of materials at the surface and internal dynamics of terrestrial planets give rise to (solid) topography, which interacts with the (liquid) hydrosphere, on Earth, and (gas) atmosphere to produce landscapes of remarkable variety.

Conventionally, geodynamics is founded on the theory of self-gravitating viscous fluids (e.g. see \cite{Bercovici2000} and references therein).  From a distant perspective this makes sense given that all planets, even the terrestrial ones, are spheroidally shaped.  The viscosity of Earth's mantle, estimated based on observations of post-glacial isostatic rebound, is generally agreed to be strongly temperature dependent and huge, something like $10^{21} Pa-s$.  Of course, this does not mean that geodynamicists actually believe that rocks are viscous fluids, but simply that rocks behave that way under certain circumstances.  What then are the limits of this geodynamic approximation?

\begin{figure} \label{Figure_1}
\centering
\includegraphics[width=0.5\textwidth]{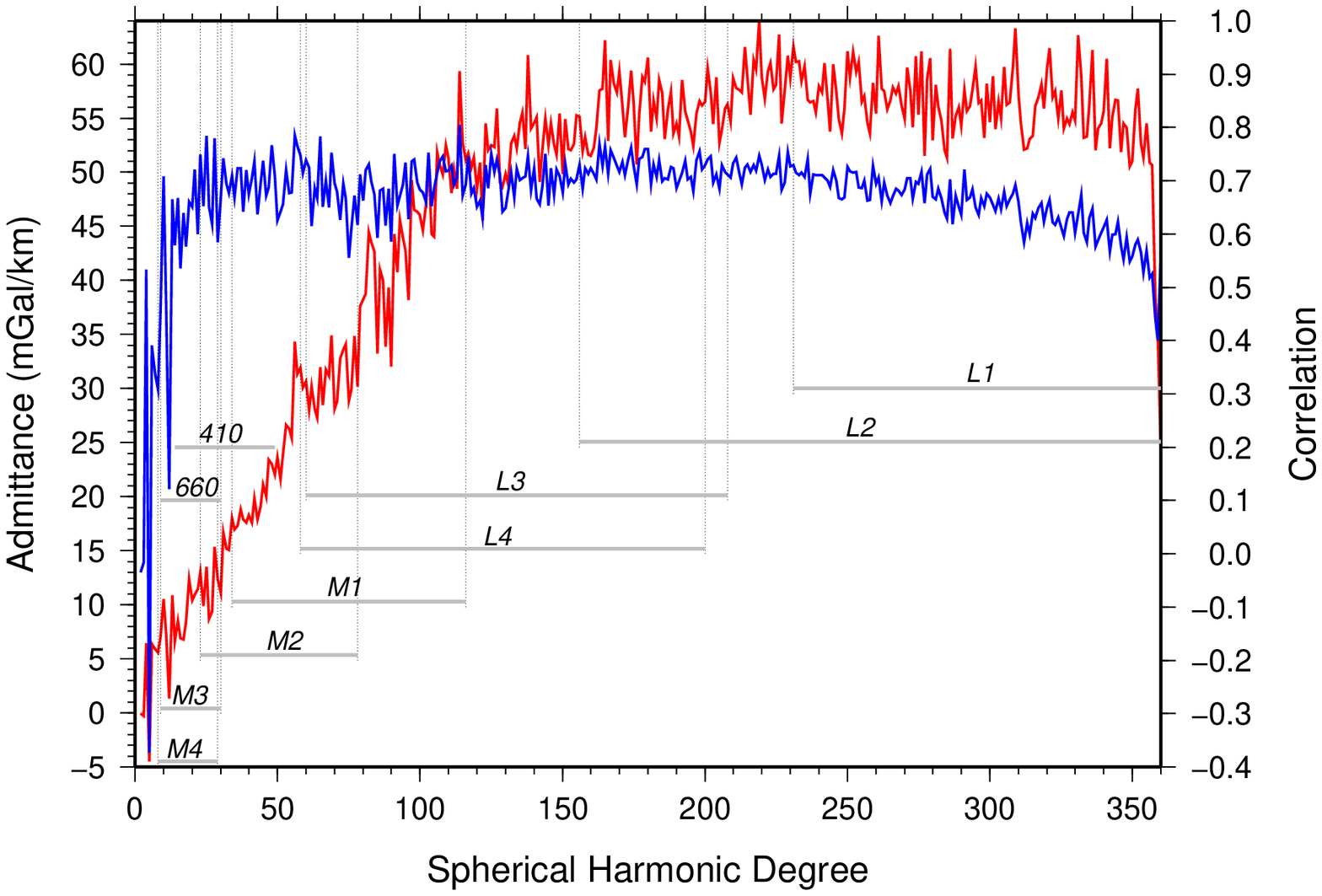}
\caption{Earth's radial gravity-topography admittance (red) and correlation (blue) spectra (after \protect{\cite{Wieczorek2007}}), compared with hypothetical wavebands for lateral strength and density heterogeneities, based on ThERM \protect{\cite{Patton2010}}.  Horizontal gray bars span from $2$ to $\zeta$ times the depths to lithospheric $L1-4$ and mantle $M1-4$ modes of that model.  Also indicated are comparable wavebands for the 410-km and 660-km seismic discontinuities \protect{\cite{Shearer2000}}.}
\end{figure}

Gravity and topography \cite{Wieczorek2007} provide important observational constraints for planetary geodynamics.  Seismicity too has proven indispensable for illuminating Earth's deeper structure \cite{Dziewonski1981} and delineating the boundaries and relative motions of its tectonic plates \cite{Isacks1968}.  Recent studies of seismic source processes using observations of strong ground motions in the near field of crustal faults suggest that some ruptures propagate at supershear speeds.  What can we infer about seismic ruptures more generally?

Geologically, Earth's crust retains a record of structural, thermal, and chemical changes spanning at least four billion years, and results for other rocky bodies are no less varied.  Consequently, an understanding of the evolution of terrestrial systems has practical applications to resource prospecting, climate change assessment, and seismic hazard mitigation, in addition to its fundamental scientific interest.

It is convenient to decompose observed surface motions into poloidal and toroidal components (e.g. \cite{Bercovici2000}).  The poloidal field captures convective motion, with its upwelling and downwelling mass transport, and associated divergence and convergence at the upper and lower boundaries of the mantle.  The toroidal field, on the other hand, captures horizontal rotational motion associated with the relative movement and spin of tectonic plates, and the strike-slip faulting which facilitates these motions \cite{Bercovici1995}.  Note that some concept of gravitation is necessary, not only for the descriptives vertical/horizontal and upper/lower to make sense, but also to explain the forces driving these motions.

Nevertheless, the dynamic and historical interpretation of geological data requires us to make simplifying assumptions about material behavior, over broad ranges of temperature, pressure, and time.  Here, a general approach to the poloidal-toroidal coupling problem, inclusive of viscous fluid behavior and featuring finite normal stress differences, is developed.  Employing the stress-deformation relation for differential grade-2 (DG-2) materials \cite{Rivlin1955,Dunn1995,Truesdell2004}, this work offers new insights for the interpretation of planetary survey data.

\section{The Interior Problem}
\subsection{Spacetime}
The everyday world is intrinsically four-dimensional, compounded from one time-like and three space-like directions, all of which are locally orthogonal.  This is the spacetime manifold of Lorentz, in which all non-gravitational interactions of matter-energy take place, and to which the rules of special relativity apply.  Here I adopt the conventions of a directed Minkowski domain in which space-like vectors have positive squared length, time-like ones have negative squared length, and the future sense of the temporal dimension is negative \cite{Bragg1964,Misner1973}.

The interactions of matter-energy in spacetime involve the transfer of momentum which, for a given particle of mass-energy $m$, can be represented as a 4-vector $\vec{p}=m\vec{\upsilon}$, where $\vec{\upsilon}$ is the 4-velocity reported by a given observer.  If the observer is in motion with respect to the particle, then the spatial components of the 4-velocity are just the usual velocities ${\upsilon}^j={v}^j; (j=1-3)$, while if the observer is at rest with respect to the particle, ${\upsilon}^j=0; (j=1-3)$.  Note that in each case the observer and the particle are always moving through time, so that ${\upsilon}^0=-1$.  Certainly, this description is adequate for a system of discrete particles, but a more comprehensive one is needed if we wish to study the evolution of complex systems, like terrestrial planets.  Fortunately, the stress-energy tensor $T$ provides such a description, and furthermore offers a direct connection between the stress-deformation relations of continuum mechanics \cite{Truesdell2004} and modern gravitation theory \cite{Misner1973}.

\vspace{10pt}
Table 2.  Symbology \par
\begin{tabular}{@{}cll@{}} \title{symbology}
Symbol & Dimension & Description\\
\hline
$\zeta$ & $-$ & the number $4\sqrt{3}$\\
$G$ &  & curvature tensor\\
$T$ & $ML^{-1}T^{-2}$ & stress-energy tensor\\
$\vec{p}$ & $MLT^{-1}$ & momentum\\
$m$ & $M$ & mass\\
$\vec{x}$ & $L$ & position $=(x,y,z)$\\
$\vec{v}$ & $LT^{-1}$ & velocity $=(u,v,w)$\\
$\rho$ & $ML^{-3}$ & mass density\\
$\sigma$ & $ML^{-1}T^{-2}$ & stress tensor\\
$\nabla$ & $L^{-1}$ & gradient operator\\
$\nabla\cdot$ & $L^{-1}$ & divergence operator\\
$\nabla\times$ & $L^{-1}$ & curl operator\\
$\Delta$ & $L^{-2}$ & 3-D Laplacian operator\\
$\Delta_h$ & $L^{-2}$ & 2-D Laplacian operator\\
$p$ & $ML^{-1}T^{-2}$ & pressure $=\smash{\frac{1}{3}{\sigma}^{kk}}$\\
$g$ & $LT^{-2}$ & gravitational acceleration\\
$\eta$ & $ML^{-1}T^{-1}$ & dynamic viscosity\\
$\beta_1,\beta_2$ & $ML^{-1}$ & normal stress moduli\\
$D()\slash{Dt}$ & $T^{-1}$ & proper derivative, also $\dot{()}$\\
$A^{(1)}$ & $T^{-1}$ & velocity strain tensor\\
$A^{(2)}$ & $T^{-2}$ & acceleration strain tensor\\
$\vartheta$ & $T^{-1}$ & cubic dilation\\
$\Omega$ & $T^{-1}$ & vorticity\\
$|\alpha|$ & $-$ & rescaling modulus\\
$\psi$ & $$ & toroidal potential\\
$Q$ & $ML^{2}T^{-2}$ & heat (First Law)\\
$U$ & $ML^{2}T^{-2}$ & internal energy\\
$W$ & $ML^{2}T^{-2}$ & work\\
$S$ & $ML^{2}T^{-2}K^{-1}$ & entropy (Second Law)\\
$\theta$ & $K$ & absolute temperature\\
$k$ & $MLT^{-3}K^{-1}$ & thermal conductivity\\
$C_p$ & $ML^{2}T^{-2}K^{-1}$ & heat capacity\\
$\epsilon$ & $K^{-1}$ & thermal expansivity\\
\hline
\end{tabular}
\vspace{10pt}

\subsection{Gravitation}
Einstein \cite{Einstein2007} accounted for gravitation using the field equation
\begin{equation}
{G}=8\pi{T}, \label{Einstein}
\end{equation}
where the curvature of spacetime ${G}$ takes its source in a finite stress-energy ${T}$.  The latter quantity has generic components
\begin{equation} \label{stress-energy}
 {T}= \begin{bmatrix}
 \rho  &  \rho{v}^k\\
 \rho{v}^j & {\sigma}^{jk}
 \end{bmatrix}
 ;j,k=1-3.
\end{equation}
where $\rho$ is mass-energy density, $\rho{v}^{j}$ are momentum flux densities or energy fluxes, and $\sigma^{jk}$ is stress \cite{Misner1973}.

By construction the curvature $G$ satisfies the contracted Bianchi identity.  Consequently its divergence must vanish $\nabla\cdot{G}\equiv0$.  In other words, the geometry of spacetime is conserved.  This, in turn, constrains the evolution of stress-energy to the equations of motion contained in $\nabla\cdot{T}=0$.  Expansion of this contraint on an arbitrary orthonormal basis leads, after some manipulation, to the continuity equation
\begin{equation}
\frac{D\rho}{Dt}+\rho\left(\nabla\cdot\vec{v}\right)=0 \label{continuity}
\end{equation}
and the balance of linear momentum
\begin{equation}
 \rho\frac{D\vec{v}}{Dt}=\nabla\cdot\sigma \label{mom_bal}
\end{equation}
familiar from continuum mechanics \cite{Truesdell2004}.  Note that ${D}\slash{Dt}$ denotes a proper derivative, evaluated in the rest frame of a given particle.  While \eqref{continuity} and \eqref{mom_bal} can be expected to apply in most circumstances, their use in a given physical problem requires further specification for $\rho$ and $\sigma$.

\subsection{Mass-energy density}
The mass-energy density shall be assumed to fall in the range of normal rocky or iron-rich matter found in meteorites.  As such, the model under construction has naught to do with gas giants or processes of atmospheric circulation.  Similarly, it has naught to do with stars or processes of stellar nucleosynthesis, let alone any of the other exotica associated with general relativity.  Still, as this work demonstrates, there are useful, if ordinary, things to be found via Einstein's approach, provided one is willing to pursue the logical development of ideas.  Key to this approach is the selection of a relationship between stress and deformation that is general enough to make predictions that can be correlated with geological observations, simple enough to allow analytical manipulation, and inclusive of other simpler relations.  A variety of kinematic elements are needed to express these relationships clearly.

\subsection{Kinematic elements}
All of the kinematic elements required for this work can be derived from the general velocity gradients tensor $\nabla{v}$, which has as many as nine unique components $\smash{\frac{\partial{v}^{i}}{\partial{x}^{j}}};(i,j=1-3)$.  As is well-known, any general second rank tensor can be decomposed uniquely into symmetric and anti-symmetric parts.  Furthermore, the symmetric part can be decomposed into trace and trace-free parts \cite{Misner1973}.  The symmetric part $A^{(1)}$, called the velocity-strain tensor, encompasses both volume and shape changes and is constructed in the usual way ${A_{ij}^{(1)}}=\smash{\frac{\partial{v}^{i}}{\partial{x}^{j}}}+\smash{\frac{\partial{v}^{j}}{\partial{x}^{i}}}$.  It has no more than six unique components.  Volume changes alone are associated with the cubic dilation $\vartheta=\nabla\cdot\vec{v}=\smash{\frac{\partial{v}^{k}}{\partial{x}^{k}}}$.  The repeated index $k$ here indicates summation over three spatial directions, consistent with the summation convention.  The anti-symmetric part $\Omega$, called the vorticity tensor, represents rotation and is constructed in the usual way ${\Omega}_{ij}=\smash{\frac{\partial{v}^{i}}{\partial{x}^{j}}}-\smash{\frac{\partial{v}^{j}}{\partial{x}^{i}}}$.  It has no more than three unique components.

\subsection{Generalized incompressibility}
Material incompressibility is another contraint commonly applied in these problems.  Usually this amounts to a steady-state version of the continuity equation \eqref{continuity}, $\vartheta=0$, where the proper derivative of the mass density is assumed to vanish.  However, because temporal changes in density must attend convection in the mantle, we shall instead adopt a generalized incompressibility contraint (e.g. \cite{Passerini2005}), where the cubic dilation and the proper derivative of temperature, here denoted by an overdot, are related by $\vartheta=\epsilon\dot{\theta}$ and $\epsilon$ is thermal expansivity.  Combined with equation \eqref{continuity}, this leads to
\begin{equation}
{\rho}={{\rho}_r}e^{-\epsilon\left(\theta-{\theta}_r\right)} \label{gen_incomp}
\end{equation}
where the subscript $r$ indicates known values of density and temperature in a convenient reference state.

\subsection{Stress-deformation relation}
The stress-deformation relation for a DG-2 material is given by
\begin{equation}
 \sigma=\eta{A^{(1)}}+\beta_1{A^{(2)}}+\beta_2{{A^{(1)}}\cdot{A^{(1)}}} \label{stress_def}
\end{equation}
where $\eta$ is dynamic viscosity, and $\beta_1$ and $\beta_2$ are the first and second normal stress moduli.  Here $A^{(1)}$ is the velocity-strain tensor, defined earlier, while the tensor $A^{(2)}$, called the acceleration-strain, has components defined by
\begin{equation}
 {A_{ij}^{(2)}}={\frac{D{A_{ij}^{(1)}}}{Dt}}+{A_{ik}^{(1)}}\cdot{\nabla{v}}_{kj}+{\nabla{v}}_{ki}\cdot{{A_{kj}^{(1)}}}. \label{accel_strain}
\end{equation}
This relation was first derived by Rivlin \& Ericksen \cite{Rivlin1955}.

Based on the work of Dunn \& Rajagopal \cite{Dunn1995} it is expected that normal stress effects proportional to $\beta_1$ and $\beta_2$ will be of the same order of magnitude.  The former is associated with finite material strength \cite{Patton2005,Patton2010,Patton2013}, and the latter with attenuation \cite{Truesdell1964}.

\subsection{Pressure}
Pressure in this model is not given by a thermodynamic equation of state.  Rather it is a mechanical pressure $p=\smash{\frac{1}{3}\sigma^{kk}}$ obtained by averaging the components appearing on the diagonal of the stress tensor.  Note that $\sigma^{kk}$ denotes the sum of these components, consistent with the summation convention.  Combining \eqref{gen_incomp} with the hydrostatic relation $p=\rho{g}{z}$, where $g$ is gravitational acceleration and $z$ is depth beneath the surface of the planet, we obtain
\begin{equation}
p={\rho_r}{g}{z}e^{-\epsilon\left(\theta-{\theta}_r\right)} \label{static_pressure}
\end{equation}
Thus, thermal variations in density perturb the static pressure and thereby drive poloidal motion.  How this motion might be coupled to toroidal motion is examined in the next section. 

\section{Quasi-static evolution equations}
According to the encyclopedia of mathematics \cite{Myshkis2013} an evolution equation is a differential law that can be interpreted as governing the development in time of a system.  This section outlines the derivation of such equations for the poloidal and toroidal fields of a DG-2 material.  The velocity potentials are extracted by taking successive curls and normal components of the balance of linear momentum \eqref{mom_bal} \cite{Boronski2006}.  Residual terms are then potential drivers for the targeted field.  Results for viscous fluids \cite{Bercovici2000}, defined by \eqref{stress_def} with $\beta_1=\beta_2=0$, in the quasi-static limit and therefore lacking explicit time dependence, are recovered as special cases.  In contrast, the results reported here for DG-2 materials exhibit explicit dependence on the natural time $\smash{\frac{\beta_1}{\eta}}$ \cite{Truesdell1964}, even in the quasi-static limit.  

\subsection{Poloidal field}
The quasi-static evolution equation for the poloidal field can be found by taking
\begin{equation}
 \vec{e}_{k}\cdot\nabla\times\nabla\times\left(\nabla\cdot\sigma\right)=0
\end{equation}
The explicit result is rather lengthy and therefore not reproduced here.  Implicitly, this relation has the form
\begin{equation}
\left(\eta+\beta_1\frac{D}{Dt}\right)\Delta\Delta_{h}w=\beta_1\Pi\left(\nabla{v},A^{(1)},\vartheta\right) \label{pol_DG-2}
\end{equation}
where $w$ is the vertical velocity component, $\Delta$ is the 3-D Laplacian, $\Delta_{h}$ is the 2-D horizontal Laplacian \cite{Boronski2006}, and the other symbols retain their meanings from above.  The residual forcing terms are non-linear, quadratic in derivatives of the velocity field, and involve first-, second-, and third-order gradients of the viscosity and normal stress moduli.  For reasons presented below in connection with the toroidal field, these material gradient terms are suppressed here.  The remaining terms $\Pi$, proportional to $\beta_1$, can be organized by the nine velocity gradient components, and the eighteen next higher spatial derivatives of them.  Note the presence of cubic dilation, and the absence of vorticity.  Very little algebraic collapse occurs in this expression, and what does can be summarized as follows.  

Each of the nine velocity gradient components appears as a cofactor for a sum of fourth-order mixed partial derivatives of the velocity components.  An obvious pattern is factors of the form $-\Delta\Delta_{h}u$, $-\Delta\Delta_{h}v$, and $-\Delta\Delta_{h}w$ multiplying gradients of the vertical velocity $\smash{\frac{\partial{w}}{\partial{x}}}$, $\smash{\frac{\partial{w}}{\partial{y}}}$ and $\smash{\frac{\partial{w}}{\partial{z}}}$, respectively.  These terms likely are associated with the convergence and divergence of material at the upper and lower boundaries of the convecting mantle.

Similarly, each of the eighteen second-order velocity derivatives appears as a cofactor for sums of mixed third-order partial derivatives of the velocity components.  Of these, only $\smash{\frac{\partial^2{u}}{\partial{z}^2}}$ and $\smash{\frac{\partial^2{v}}{\partial{z}^2}}$ multiply factors of the form $\smash{\frac{\partial^2{\vartheta}}{\partial{x}\partial{z}}}$ and $\smash{\frac{\partial^2{\vartheta}}{\partial{y}\partial{z}}}$.  These terms, involving the cubic dilation, provide coupling to thermal density changes, and thereby give rise to non-isotropic pressure gradients in the mantle.

In the absence of normal stress differences \eqref{pol_DG-2} reduces to
\begin{equation}
\eta\Delta\Delta_{h}w=0. \label{pol_Newtonian}
\end{equation}
This result differs from that of Bercovici \emph{et al} \cite{Bercovici2000}, because they assumed an explicit body force term for gravity in \eqref{mom_bal}, consistent with the Boussinesq approximation, whereas here the effects of gravity have been subsumed into the field equation \eqref{Einstein}.  Consequently, poloidal motions in this model are driven by pressure gradients, not buoyancy forces.

\subsection{Toroidal field}
The quasi-static evolution equation for the toroidal field can be obtained by taking
\begin{equation}
 \vec{e}_{k}\cdot\nabla\times\left(\nabla\cdot\sigma\right)=0
\end{equation}
Again, the complete result is rather lengthy and therefore omitted here.  First- and second-order gradients of the viscosity and normal stress coefficients appear explicitly, but only the first-order gradients are exhibited here in the general relation
\begin{multline}
\label{tor_evol}
\Delta\Delta_{h}\psi+{\frac{\beta_1}{\eta}}\Bigg\{{\frac{D}{Dt}}\left(\Delta\Delta_{h}\psi\right) \\
+{\frac{\partial^2\psi}{\partial{x}\partial{y}}}\left(2{\frac{\partial^4\psi}{\partial{y}^4}}-2{\frac{\partial^4\psi}{\partial{x}^4}}+3{\frac{\partial^4\psi}{\partial{y}^2\partial{z}^2}}+3{\frac{\partial^4\psi}{\partial{x}^2\partial{z}^2}}\right)\\
+\left({\frac{\partial^2\psi}{\partial{x}^2}}-{\frac{\partial^2\psi}{\partial{y}^2}}\right)\left(2{\frac{\partial^4\psi}{\partial{x}^3\partial{y}}}+2{\frac{\partial^4\psi}{\partial{x}\partial{y}^3}}+3{\frac{\partial^4\psi}{\partial{x}\partial{y}\partial{z}^2}}\right)\Bigg\}\\
 ={\frac{1}{\eta}}\Theta\left(\nabla{v},A^{\left(1\right)},\Omega;\beta_1,\beta_2,\nabla\eta,\nabla\beta_1,\nabla\beta_2\right)
\end{multline}
Here $\psi$ is the toroidal potential, and the other symbols retain their meanings from above.  The residual forcing terms $\Theta$ are non-linear, quadratic in derivatives of the velocity field.  Note the presence of the vorticity, and the absence of the cubic dilation.

In the absence of normal stress differences equation \eqref{tor_evol} reduces to
\begin{equation}
 \Delta\Delta_{h}\psi=\frac{1}{\eta}\Theta\left(A^{(1)};\nabla\eta\right) \label{tor_visc}
\end{equation}
While all of the velocity-strain components, except for ${A_{33}}^{(1)}$, appear in this expression, the vorticity components implicit in \eqref{tor_evol} do not.  Additionally, if the viscosity gradients were taken to vanish, then the right-hand side of \eqref{tor_evol} would also vanish.  Thus, in conventional geodynamic models, viscosity gradients are needed to drive toroidal motions \cite{Bercovici1995,Bercovici2000}.

Further analysis, however, shows this to be neither necessary nor sufficient in the presence of finite normal stress differences.  Upon neglecting \emph{all} material gradient terms, equation \eqref{tor_evol} reduces to
\begin{multline}
 \Delta\Delta_{h}\psi+\frac{\beta_1}{\eta}\Bigg\{{\frac{D}{Dt}}\left(\Delta\Delta_{h}\psi\right)\\
 +\frac{\partial^2\psi}{\partial{x}\partial{y}}\left(2\frac{\partial^4\psi}{\partial{y}^4}-2\frac{\partial^4\psi}{\partial{x}^4}+3\frac{\partial^4\psi}{\partial{y}^2\partial{z}^2}+3\frac{\partial^4\psi}{\partial{x}^2\partial{z}^2}\right)\\
 +\left(\frac{\partial^2\psi}{\partial{x}^2}-\frac{\partial^2\psi}{\partial{y}^2}\right)\left(2\frac{\partial^4\psi}{\partial{x}^3\partial{y}}+2\frac{\partial^4\psi}{\partial{x}\partial{y}^3}+3\frac{\partial^4\psi}{\partial{x}\partial{y}\partial{z}^2}\right) \Bigg\}\\
 =\frac{1}{\eta}\Theta\left(\nabla{v},A^{(1)},\Omega;\beta_1\right). \label{tor_DG-2}
 \end{multline}
Note that vorticity terms are retained, consistent with the documented dynamic connection between normal stress differences and vorticity \cite{Patton2010}. Therefore residual terms proportional to $\beta_1$ can drive toroidal motions.  Viscosity gradients are not required.  Moreover, it appears that material gradients of any kind are not required.  This suggests that a useful simplification of this non-linear system can be achieved by neglecting these terms.

In any case material gradients lack intrinsic physical meaning, except in relation to a particular configuration \cite{Truesdell2004}.  This raises the issue of what defines a reference configuration.  Certainly, the gross spherical symmetry of terrestrial planets is one important aspect, which is explained neatly by Schwarzschild's (ca. 1916) exact solution of the field equation \eqref{Einstein}, and spheroidal extensions of that metric by Kerr and Newman \cite{Misner1973}.  But how can we reconcile the ideal "stress-free" reference state of Hookean elasticity, and the notion of hydrostatic equilibrium required by the definition of a planet, with the pervasive and irreversible deformation evident in crustal and mantle rocks?

\section{A geodynamic approximation}
The viscosity of Earth's mantle, consistent with observations of post-glacial isostatic rebound, is generally agreed to be huge, something like $10^{21} Pa-s$ \cite{Bercovici2000}.  This, in fact, is what motivates neglect of the acceleration terms on the left-hand side of \eqref{mom_bal} - the quasi-static approximation.  If $\beta_1$ is assumed to be finite, then we can approximate \eqref{tor_DG-2} as
\begin{multline}
 \Delta\Delta_{h}\psi+\frac{\beta_1}{\eta}\Bigg\{\frac{D}{Dt}\left(\Delta\Delta_{h}\psi\right)\\
 +\frac{\partial^2\psi}{\partial{x}\partial{y}}\left(2\frac{\partial^4\psi}{\partial{y}^4}-2\frac{\partial^4\psi}{\partial{x}^4}+3\frac{\partial^4\psi}{\partial{y}^2\partial{z}^2}+3\frac{\partial^4\psi}{\partial{x}^2\partial{z}^2}\right)\\
 +\left(\frac{\partial^2\psi}{\partial{x}^2}-\frac{\partial^2\psi}{\partial{y}^2}\right)\left(2\frac{\partial^4\psi}{\partial{x}^3\partial{y}}+2\frac{\partial^4\psi}{\partial{x}\partial{y}^3}+3\frac{\partial^4\psi}{\partial{x}\partial{y}\partial{z}^2}\right) \Bigg\}\\
 =0
\end{multline}
Using linear perturbations about a pure-shearing plane strain, 2-D, base velocity field, this expression reduces to a fifth order partial differential equation
\begin{multline}
\left(1-2{\frac{\beta_1}{\eta}}\right){\frac{\partial^4{\psi}}{\partial{x}^4}}+2{\frac{\partial^4{\psi}}{\partial{x}^2\partial{y}^2}}+\left(1+2{\frac{\beta_1}{\eta}}\right){\frac{\partial^4{\psi}}{\partial{y}^4}}\\
=-{\frac{\beta_1}{\eta}}{\frac{D}{Dt}}\Delta\Delta_{h}\psi
\end{multline}
Upon dividing through by $1+2\beta_1\slash\eta$, and defining the function
\begin{equation}
 {\alpha^2}={\frac{1-2\beta_1\slash\eta}{1+2\beta_1\slash\eta}} \label{rescaling}
\end{equation}
we obtain an inhomogeneous diharmonic equation
\begin{multline}
 {\alpha^2\frac{\partial^4{\psi}}{\partial{x}^4}}+{\left(1+\alpha^2\right)\frac{\partial^4{\psi}}{\partial{x}^2\partial{y}^2}}+{\frac{\partial^4{\psi}}{\partial{y}^4}}\\
 = -{\frac{\beta_1}{\left(\eta+2\beta_1\right)}}{\frac{D}{Dt}}\Delta\Delta_{h}\psi\label{diharmonic}
\end{multline}
This suggests that the evolution of toroidal motions should exhibit diharmonic scaling properties.  The relevance of this result to terrestrial dynamics, and the interpretation of gravity, topography, and seismicity data, requires an understanding of the thermodynamics of DG-2 materials.

\section{Thermodynamics}
\subsection{Preliminaries}
The macroscopic notion of heat is defined as the difference between the internal energy and work performed on a system, consistent with the Joule heating experiments \cite{Chandrasekhar1967}.  The First Law of thermodynamics is therefore $dQ=dU-dW$, where $dQ$, $dU$ and $dW$ are increments of heat, internal energy and work, respectively.  Note that heat is a derived quantity, having no meaning independent of the First Law.

The Second Law of thermodynamics was formulated axiomatically by Carathéodory (ca. 1909) based on an analysis of Pfaffian differential equations to read $dS={dQ}\slash\theta$, where $dS$ is an increment of the entropy density, and $\theta$ is absolute temperature. Thanks to Boltzmann (ca. 1877) and Planck (ca. 1920), this relation provides a bridge between macroscopic and microscopic physics.

Guided by rock mechanics experiments, we can begin accounting for the energy and entropy densities of strained solid materials using a simple one-dimensional elastic model.  This model exhibits an unorthodox behavior consistent with Lavenda's notion of thermodynamic symmetry breaking \cite{Lavenda1995}.  The expected slope, shape, and temperature dependence of the energy density for this model serves as foil for the shear localization mechanism of DG-2 materials \cite{Patton2010,Patton2013}.

\subsection{Strained Hookean solids}
Consider a cylindrical test specimen of rock, with length $l$ and diameter $d$, placed in a loading frame for the purpose of strength characterization.  In response to a force $\phi$ directed along a line parallel to the specimen's length, the specimen shortens by a length increment $dl$.  Consequently, the increment of work needed to shorten the cylinder from $l+dl$ to $l$ is given by ${dW}=\phi{dl}$.  Because the specimen can be held under relatively small loads for long periods of time, it is reasonable to assume that it manifests a force equal and opposite to the applied load.  Presumably, this reaction force arises from electromagnetic interactions in the sample's microstructure.  This is Hooke's law \emph{Ut tensio sic vis} (ca. 1642).

However, it is equally valid to consider this problem from an energetic point of view.  Responding to a directed environmental load of magnitude $\phi$, the cylinder strains by an increment $dl$ of its overall length $l$, and as a result distributes an increment of energy $dU$ throughout its microstructure and mineral fabric.  Therefore, we have a macroscopic relation for the energy increase of the form ${dU}=\phi{dl}$.

Eliminating $dQ$ between the First and Second Laws leads to the equilibrium relation
\begin{equation}
\theta{dS}=dU-dW.  \label{therm_sym}
\end{equation}
Upon substituting the work and internal energy increments from above we find, for positive absolute temperatures, that $dS=0$.  Thus for strained Hookean solids, thermodynamics predicts no change in entropy, consistent with the apparent lack of energy dissipation, and a state of mechanical equilibrium for cylinders under small loads.

Curiously, and despite its fundamental place in thermodynamics, there is no need to account for heat in these experiments.  Therefore, no meaningful distinction can be made between heat and work for this model \cite{Lavenda1995}.  The potentials for work, internal energy, and entropy are all inhomogeneous functions of a single variable, rather than the homogeneous functions assumed at equilibrium.  This has consequences for the assumed combination of the First and Second Laws.

\subsection{Strained inhomogeneous solids}
Experience shows that if the cylinder above were subsequently unloaded, it would likely return to its original length.  The unloaded state therefore is somehow more likely than the loaded one, and should coincide with a local maximum in entropy.  Consequently, any deformation of the cylinder from this ideal reference state must necessarily decrease the entropy of the cylinder itself, as a function of length.

If heat and work are indistinguishable, and heat is accounted for by the product of the temperature and the entropy increment through the Second Law, \eqref{therm_sym} can be rewritten as
\begin{equation}
{\theta}=-\frac{d\delta{U}}{d\delta{S}}.  \label{non_sym}
\end{equation}
Here the entropy and internal energy are prefixed with deltas to distinguish these primitive functions from the homogeneous thermodynamic potentials assumed above.  These functions can manifest scale--dependence, contrary to the scalability expected from classical thermodynamics.

The importance of variability in the behavior of elastic materials, and rocks in particular, can be demonstrated by modeling the energy and entropy potentials for this system as power laws in length $l$.  We define the internal energy increase as
\begin{equation}
 \delta{U(l)}={\frac{\lambda}{m}}{l}^{m} \label{energy_increase}
\end{equation}
and the entropy reduction as
\begin{equation}
 \delta{S(l)}=-{\frac{k\gamma}{n}}{l}^{n} \label{entropy_reduction}
\end{equation}
where $\lambda$ and $\gamma$ are positive constants independent of temperature, $m$ and $n$ are positive numbers, and $k$ is Boltzmann's constant.  Substituting derivatives of \eqref{energy_increase} and \eqref{entropy_reduction} into equation \eqref{non_sym} we obtain
\begin{equation}
 \theta={\frac{\lambda}{k\gamma}}{l}^{m-n}. \label{temperature}
\end{equation}
Consequently, the temperature of this model system can either increase or decrease with length, depending on whether the exponent $m-n$ is positive or negative.  The temperature is independent of length for $m = n$.

The elasticity modulus $E$ for this model is defined by the derivative of the force $\phi$, which in turn is the derivative of the internal energy $\phi=\smash{\frac{dU}{dl}}$.  Upon eliminating length in this expression via the temperature relation \eqref{temperature}, we find  
\begin{equation}
 E=(m-1)\lambda({\frac{k\gamma{\theta}}{\lambda}})^{\frac{m-2}{m-n}} \label{elasticity_modulus}
\end{equation}
For $n = 2$, the modulus is $E=(m-1)k\gamma{\theta}$, and the force reduces to a generalized Hooke's law $\phi=E(\theta)l$.  This relation further reduces to a linear force-displacement law, but only when the internal energy too is quadratic in length, $m = 2$.

Upon inverting equation \eqref{temperature} to express length as a function of temperature, differentiating the result with respect to temperature, and eliminating the constants via \eqref{temperature} we obtain
\begin{equation}
 {\frac{1}{l}}{\frac{dl}{d\theta}}={\frac{1}{(m-n)\theta}}. \label{thermal_elongation}
\end{equation}
This expression characterizes the thermal elongation of the model.  Consequently the model elongates upon heating for $m > n$, shortens upon heating for $m < n$, and is undefined for $m = n$.  This is analogous to the thermal expansivity $\epsilon$ in three-dimensions, which dictates the magnitude of thermal variations in mass density.

By separating the entropy and energy increments in equation \eqref{non_sym}, dividing through by a temperature increment $d\theta$, expressing the common length dependencies using the chain rule, and finally substituting derivatives with respect to length of equations \eqref{energy_increase}, \eqref{entropy_reduction}, and \eqref{temperature}, we find
\begin{equation}
 {\frac{d\delta{U}}{d\theta}}=-\theta{\frac{d\delta{S}}{d\theta}}=({\frac{\lambda}{m-n}}){\frac{{l}^{m}}{\theta}}. \label{broken_symmetry}
\end{equation}
For positive absolute temperature, \eqref{broken_symmetry} shows that the heat capacity of this inhomogeneous elastic system cannot be defined simultaneously as
\begin{equation}
 C\equiv{\frac{dQ}{d\theta}}=\theta{\frac{d\delta{S}}{d\theta}} \label{entropy_density}
\end{equation}
and
\begin{equation}
 C\equiv{\frac{dQ}{d\theta}}={\frac{d\delta{U}}{d\theta}} \label{energy_density}
\end{equation}
because one of these definitions always will be negative when the other is positive, and vice versa.  Apart from the pathological case for $m = n$, there are two other distinct types of inhomogeneous elastic systems depending on whether $m < n$ or $m > n$.

The variability of the energy and entropy as functions of length is inversely proportional to the exponent appearing in the primitive power laws.  Hence, a smaller exponent means greater variability.  The internal energy increase \eqref{energy_increase} clearly is associated with the macroscopic mechanical properties of the system.  On the other hand, because heat is not evident in this problem, the entropy reduction \eqref{entropy_reduction} must be associated with the microscopic statistical properties of the system, rather than its macroscopic thermal properties.  Consequently, systems dominated by variability in either mechanical $m < n$ or statistical $m > n$ properties can be identified on the basis of their heat capacity, \eqref{entropy_density} or \eqref{energy_density}, respectively.

These conclusions can be clarified by examining the relationship between energy and entropy.  Eliminating length between equations \eqref{energy_increase} and \eqref{entropy_reduction}, we find for the case of mechanical variability that
\begin{equation}
  \delta{S}\sim-(\delta{U})^{\frac{n}{m}} \label{concave_energy}
\end{equation}
In words, the entropy must be a concave function of the internal energy.  Similarly for the case of statistical variability we find
\begin{equation}
 \delta{U}\sim{(|\delta{S]|)}}^{\frac{m}{n}}. \label{convex}
\end{equation}
In words, the internal energy must be a convex function of the entropy.  Furthermore, because $\delta{dS}\slash\delta{dU} < 0$, these representations are mutually exclusive.  The usual symmetry of the entropy and energy representations, expected from equilibrium thermodynamics and arising from first-order homogeneity of thermodynamic potentials, is broken \cite{Lavenda1995}.

\begin{figure} \label{Figure_2}
\centering
\includegraphics[width=0.5\textwidth]{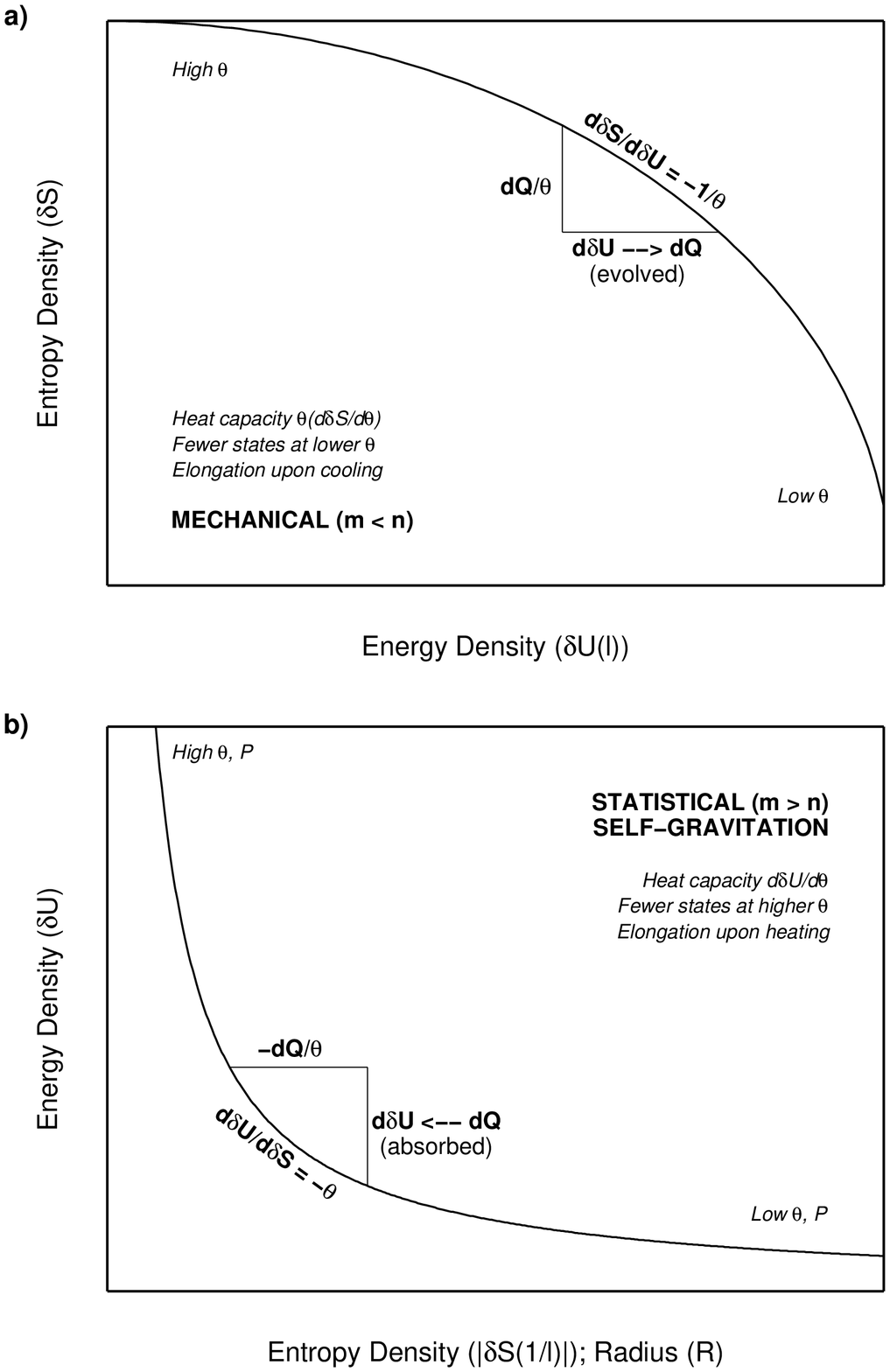}
\caption{Summary of thermodynamic relations for inhomogenous elastic solids dominated by: a) mechanical variability, b) statistical variability and/or self-gravitation (after \protect{\cite{Lavenda1995,Patton2013}}).}
\end{figure}

For an inhomogeneous elastic system dominated by mechanical variability $(m < n)$, $dQ$ is the amount of heat evolved by the system, which leads to a decrease in entropy by an amount $dS=dQ/\theta$ (Fig. 2a).  Therefore fewer microscopic states are available at lower temperatures.  The slope of the concave entropy density function is $-1/\theta$; higher temperatures are associated with flatter slopes, and lower temperatures with steeper slopes.  This model elongates upon cooling.  Because temperature and heat are both decreasing functions of length, the entropy density is proportional to length.  Mechanical variability offers no insight for the thermodynamics of DG-2 materials. 

On the other hand, in an inhomogeneous elastic system dominated by statistical variability $(m > n)$, $dQ$ is the amount of heat absorbed by the system, which leads to an increase in internal energy by an amount $dU = dQ > 0$ (Fig. 2b).  This corresponds to a decrease in the entropy by an amount $dS = -dQ/\theta$.  Consequently there are fewer microscopic states available at higher temperatures.  The slope of the convex energy density function is $-\theta$; higher temperatures are associated with steeper slopes, and lower temperatures with flatter ones.  This model elongates upon heating.  Furthermore, because temperature and heat are both increasing functions of length, energy density is inversely proportional to length.  Statistical variability offers crucial insights for the thermodynamics of DG-2 materials.

\subsection{Heat transfer and non-dimensionalization}
Now that temperature has been introduced, we must account for heat transfer in our model.  For simplicity, assume this takes place via a convection-diffusion equation for temperature given by
\begin{equation}
\rho{C_{p}}\left(\frac{\partial\theta}{\partial{t}}+\vec{v}\cdot\nabla\theta\right)=\nabla\cdot\left({k}\nabla\theta\right). \label{convect-diffuse}
\end{equation}
Here $k$ is thermal conductivity, consistent with Fourier conduction (ca. 1822), $C_{p}$ is heat capacity, and the remaining symbols retain their meanings from above.  The material constants can be collected into the thermal diffusivity $\smash{\kappa=\frac{k}{\rho{C_{p}}}}$.  Non-dimensionalization \cite{Patton2010} then allows us to express the rescaling function \eqref{rescaling} in terms of thermomechanical competence $\smash{\frac{\kappa}{\chi}}$, viz.
\begin{equation}
 {\alpha^2}={\frac{1-2\kappa\slash\chi}{1+2\kappa\slash\chi}},  \label{rescaling_non-dim}
\end{equation}
where the natural time $\smash{\frac{\beta_1}{\eta}}$ has been subsumed into a diffusivity $\chi=\smash{\frac{\eta{l^2}}{\beta_1}}$.  The latter quantity represents the rate at which dislocations diffuse through an inhomogeneous solid material, resulting in irreversible deformation.

\begin{figure} \label{Figure_3}
\centering
\includegraphics[width=0.5\textwidth]{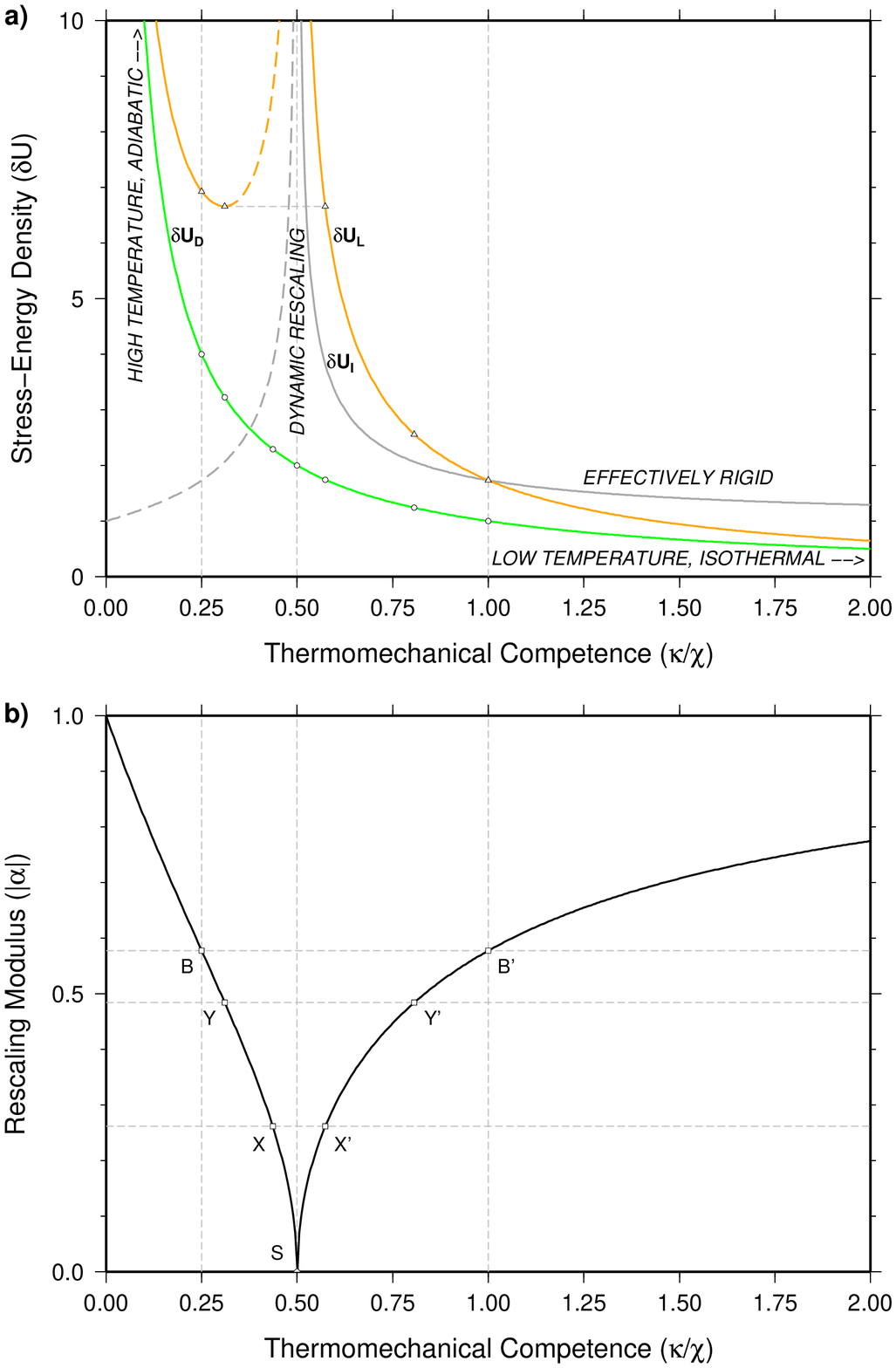}
\caption{Thermodynamic relations for DG-2 materials: a) energy thresholds for distributed ${\delta}U_D$, intrinsic ${\delta}U_I$, and localized ${\delta}U_L$ deformation; b) rescaling modulus with superposed modes (after \protect{\cite{Patton2005,Patton2010,Patton2013}}).}
\end{figure}

\subsection{Dynamic shear ruptures}
Nowhere in this simple model has the phenomenon of shear failure been addressed.  Experience shows that if we repeatedly load our test specimen or apply progressively higher loads the specimen will, at some point, spontaneously fail, sometimes after suffering significant microphysical damage \cite{Katz2004}.  In other words, the act of loading the test cylinder alters its microstructure and mineral fabric.  Although the detailed distribution of these alternations cannot be known to an outside observer, they can be treated statistically, as was appreciated by Weibull (ca. 1939).  These issues were addressed in \cite{Patton2010,Patton2013}, and are extended here to dynamic shear ruptures.

Shear localization in DG-2 materials can be represented graphically \cite{Patton2005,Patton2010} (Fig. 3a) using three energy thresholds, one for distributed harmonic deformations (green curve, ${\delta{U}_D = ({\kappa}\slash{\chi})^{-1}}$), one for intrinsic strain-energy storage (gray curve, ${\delta{U}_I = (|\alpha|)^{-1}}$), and another for localized shearing deformations (orange curve, ${\delta{U}_L = ({|\alpha|\kappa}\slash{\chi})^{-1}}$).  Taken together, these curves define an invariant energy density map for these materials.  Observe that the three threshold curves are monotonically decreasing on certain domains of thermomechanical competence, and that the energy density approaches infinity ("blows up") at the lower end of these respective domains.  The distributed threshold curve has a vertical asymptote at $\smash{\frac{\kappa}{\chi}} = 0$, while the intrinsic curve has one at $\smash{\frac{\kappa}{\chi} = \frac{1}{2}}$.  The localization threshold curve, defined as the product of the other two, consequently exhibits two asymptotes, with a distinct non-zero energy minimum between them.  The presence of two distinct energy spikes in this diagram, and their diffusive connection via the dynamic rescaling theorem \cite{Patton2010}, give rise to all of the geologically interesting behavior of the DG-2 material.

Based on incipient modes analysis, it is known that the steady-state version of the diharmonic equation \eqref{diharmonic} admits shear dislocations for values of thermomechanical competence $\smash{\frac{\kappa}{\chi}>\frac{1}{2}}$ (Fig.4a, red curve) \cite{Patton2010}.  By factoring the operator, and defining the function
\begin{equation}
\Omega\equiv\left[{{\frac{\partial^2}{\partial{x}^2}}+{\frac{\partial^2}{\partial{y}^2}}}\right]\psi=\Delta_{h}\psi,
\end{equation}
this equation can also be written as
\begin{equation}
 \Re\Omega=\left[\alpha^2{{\frac{\partial^2}{\partial{x}^2}}+{\frac{\partial^2}{\partial{y}^2}}}\right]\Omega=0 . \label{wave_eqn}
\end{equation}
Equation \eqref{wave_eqn} can be interpreted as a vector wave equation, in wavefront normalized coordinates, for motions occurring in the near field of a propagating in-plane crack provided
\begin{equation}
\frac{{v_c}^2}{{V_{S}}^2}=1-{\alpha}^2 \label{SV}
\end{equation}
Here $v_c$ is the rupture speed of the crack, and $V_{S}$ is the shear wavespeed of the material.  Note that compressional waves are also generated by an in-plane dislocation \cite{Aki2002}.

In order to eliminate the stress and velocity singularities present in homogeneous dynamic rupture models, Barenblatt \cite{Barenblatt1959} introduced an anelastic cohesive zone of width $d$ at the tip of a propagating rupture.  Comparing that work with \cite{Patton2010}, we find the relation
\begin{equation}
 {\frac{1}{|\alpha|}}={\frac{\pi}{2}} \left({\frac{\mu\gamma}{{\sigma}_c^{2}{d}}}\right),
\end{equation}
where $\mu$ is shear rigidity, $\gamma$ is Griffith surface energy, and ${\sigma}_c$ is cohesive stress.  Later, Ida \cite{Ida1972} introduced slip-rate dependent cohesion, with critical slip distance $D$.  Comparing that work with \cite{Patton2010}, we find the relation 
\begin{equation}
 {\frac{1}{{|\alpha|}}}=\frac{\pi}{4}\left(\frac{\mu{D}}{{\sigma}_c{d}}\right).
\end{equation}
Observe that these dimensionless relations suggest a common relation of the form $\smash{\xi\sim\frac{d}{|\alpha|}}$, consistent with the rescaling modulus $|\alpha|$ (Fig. 3b) found in the theory of DG-2 materials \cite{Patton2005}.  Note that this rescaling is exact in the vorticity $\Omega$ \cite{Patton2010}. 

\begin{figure} \label{Figure_4}
\centering
\includegraphics[width=0.5\textwidth]{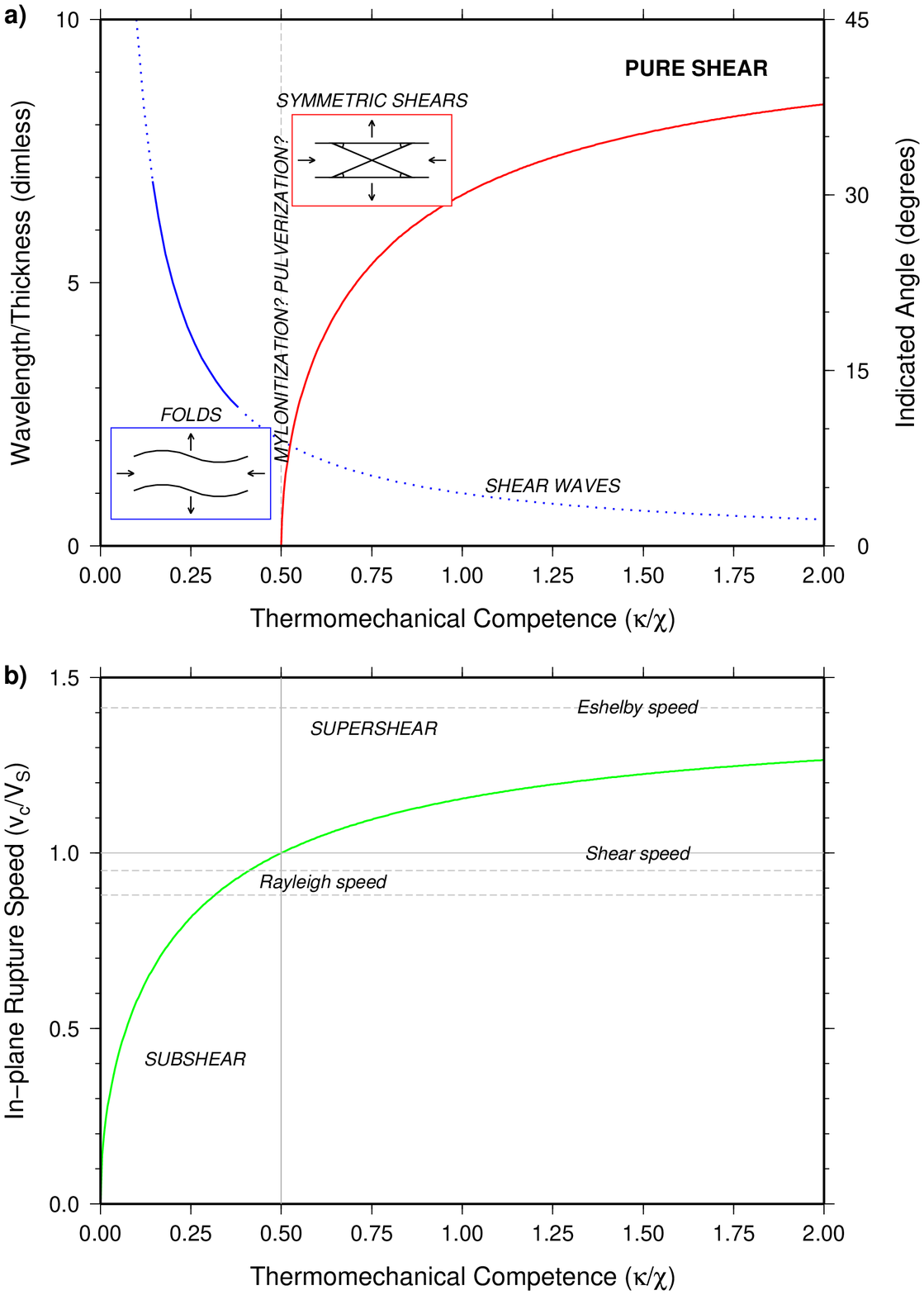}
\caption{Correlation of incipient deformation modes (after \protect{\cite{Patton2013}}) with predicted dynamic rupture speeds for DG-2 materials, as functions of thermomechanical competence: a) fold-like (blue) and shear dislocation (red) modes ; b) rupture speed for in-plane (Mode II) cracks.}
\end{figure}

Together, equations \eqref{rescaling_non-dim} and \eqref{SV} predict that crack rupture speed depends on thermomechanical competence.  Specifically, as competence increases from $0$ to $\infty$, $\alpha^2$ decreases from $1$ to $-1$, taking the value $0$ at $\smash{\frac{\kappa}{\chi}=\frac{1}{2}}$ .  Consequently, rupture speed must increase with increasing competence, from $0$, through the value $V_S$ at $\smash{\frac{\kappa}{\chi}=\frac{1}{2}}$, up to the Eshelby speed $\smash{\frac{v_c}{V_S}}\sim\sqrt{2}$.  Consequently, this model predicts supershear rupture speeds for competence greater than $\smash{\frac{1}{2}}$.  This supershear domain coincides with that for which spontaneous shear localization is possible \cite{Patton2013} (Fig. 4b).

Thus, the dynamic rescaling mechanism \cite{Patton2010} models a shear dislocation source with rupture speed $v_c$.  These findings further reinforce earlier conclusions that DG-2 materials capture the non-linear solid response of a continuum, appropriate for modeling deformation in terrestrial planets \cite{Patton2010}.

\subsection{Self--gravitating matter configurations}
Lavenda \cite{Lavenda1995} shows that the thermodynamics of a self-gravitating body, like a planet or star, is also subject to symmetry breaking of the type outlined above.  In this case, like that of inhomogeneous elastic systems dominated by statistical variability, the energy density is given by a monotonically decreasing function of the entropy density (Fig. 2b).  Significantly, the energy density is inversely proportional to the body's radius.  In other words, both pressure and temperature increase with depth.  The heat capacity of a self--gravitating body is then given by \eqref{energy_density}.

\subsection{Summary}
The energy density function for a self-gravitating, inhomogeneous elastic body, dominated by statistical variability (${m}>{n}$), must be a monotonically decreasing convex function of the radial coordinate and the entropy reduction.  Furthermore, the entropy reduction itself is an inverse function of length.  Heat absorbed by such a system will tend to increase the internal energy, but correspondingly decrease the entropy.  Such a body will not readily evolve heat, except when local conditions favor a return to more classical thermodynamics.  In these classical subsystems, the energy density function would necessarily exhibit a positive slope.

The monotonically increasing branch of the localization threshold (orange dashed curve, Figure 3a) exhibits these characteristics.  It also correlates, via ThERM \cite{Patton2009,Patton2010}, with depths in the lower crust (sub-$H3$), asthenosphere (sub-$L3$), and lower mantle (sub-$M3$) where magmas of granitic, basaltic, and komatiitic character are thought to originate.  Magmatic differentiation provides a mechanism for generating density contrasts between continental and oceanic crust, and the tectosphere and residual mantle.

The foregoing considerations are quite general and place, once and for all, the stress-energy density thresholds of DG-2 materials \cite{Patton2005} in a coherent thermodynamic context.  Consequently, the behavior of these ideal materials can be correlated with the pressure, temperature, age, and geometry of geological structures observed in outcrops, orogens, and terrestrial planets.  For example, the outer parts of such planets are predicted to be relatively cold, competent, and subject to dynamic shear localization ("brittle"), while the inner parts are predicted to be relatively hot, incompetent ("ductile"), and structurally simple.  For Earth, this is reflected in the remarkable correlation of spherically symmetric elastic models, like PREM \cite{Dziewonski1981}, with the predicted depth distribution of isobaric shears in a body with a $100 km$ thick lithosphere (Figs. 5,6) \cite{Patton2001,Patton2009,Patton2010}.  This correlation also holds for observed regional variations in earthquake depth-moment release curves.

\section{Depth-moment release curves} 
Seismic moment $M_0$, defined by the product $M_0=\mu\bar{u}(t)A$, where $\mu$ is rigidity, $\bar{u}(t)$ is average slip, and $A$ is the area of a fault, is a fundamental scalar measure of the strength of an earthquake caused by fault slip.  It is the physical basis for the moment magnitude \cite{Kanamori1977}, viz. 
\begin{equation}
M_W=\frac{2}{3}logM_0-10.7, \label{moment_mag}
\end{equation}
now routinely estimated and reported in earthquake catalogs.

If the dimensions of a slipping fault are small compared to the wavelength of radiated waves, and the duration of the slip event is short compared to the period of radiated waves, then a linear relationship exists between observed ground motions and the six independent elements of a symmetric tensor.  This latter quantity, called the seismic moment tensor, depends on source strength as well as fault orientation.  Its six elements, combined with latitude, longitude, depth, and origin time comprise the point-source centroid moment tensor (CMT) representation for an earthquake, which can be estimated for earthquakes of sufficient size, presently about $M_W=5.0$ \cite{Ekstrom2012}.

The global CMT catalog (www.globalcmt.org) includes 30,872 earthquakes recorded during the period January 1976 through December 2010 .  Of these only 20,646 were inverted automatically by the CMT algorithm and therefore have quantitative error estimates.  The consistent algorithmic treatment of these events, and the fact that a CMT estimates the anelastic work done at the source of a seismic dislocation, make them ideal for tectonic analysis. The remaining 10,266 events either had their focal depths fixed by an analyst, or were constrained by an inversion of short-period data.  

Mean standard errors for CMT latitude, longitude, and depth are $0.042^{\circ}$, $0.047^{\circ}$, and $2.82 km$, respectively.  The depth error estimate defines the minimum thickness of a filter used to smooth the earthquake depth-moment release curves presented here.  A filter smaller than this tends to display more noise than a larger one, while larger filters discard potentially interpretable depth signal in the dataset, at least at typical crustal depths.  The depth-moment release curves presented here are smoothed using either $3 km$ or $10 km$ filters.  

Earthquake depth-moment release curves $\Sigma{M}_{W}(z;t)$ represent the sum of moment magnitudes for earthquakes, filtered for depth $z$ using a boxcar of thickness $t$, at every kilometer from the surface to about $700 km$ depth.  They are an elaboration upon similar curves presented by Isacks \emph{et al} \cite{Isacks1968} in their classic paper.  In early work with the CMT catalog, filter thicknesses of $1$, $3$, $5$, $7$, $9$, and $11 km$ were used, but did not significantly affect the depth patterns shown here.  The primary effect of filter thickness, apart from curve smoothing, is to change the amplitude of the sums.

The amplitude of depth-moment release curves is well-correlated with the number of events in the given data subset.  Consequently, an expedient normalization scheme is to divide $\Sigma{M}_{W}(z;t)$ by the number of events $N$ in the subset.  Event normalized earthquake depth-moment release curves $\Sigma{M}_{W}(z;t)/N$ (Figs. 5,6) reveal significant regional differences, particularly between divergent and transform margins on the one hand and convergent margins on the other.

\begin{figure} \label{Figure_5}
\centering
\includegraphics[angle=-90,width=0.5\textwidth]{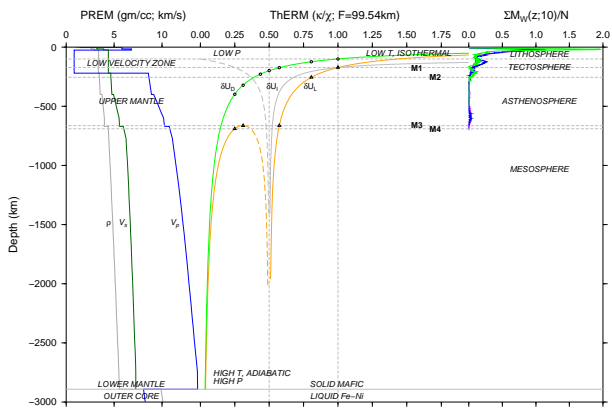}
\caption{Mantle wide correlation of:  (left) Earth's radial elastic structure \protect{\cite{Dziewonski1981}}; (center) ThERM scaled to $F=99.54km$ \protect{\cite{Patton2009,Patton2010}}; and (right) event normalized depth-moment release curves $\Sigma{M}_{W}\slash{N}$ for convergent ocean-ocean (purple), ocean-continent (blue), and continent-continent (green) margins.}
\end{figure}

\begin{figure} \label{Figure_6}
\centering
\includegraphics[angle=-90,width=0.5\textwidth]{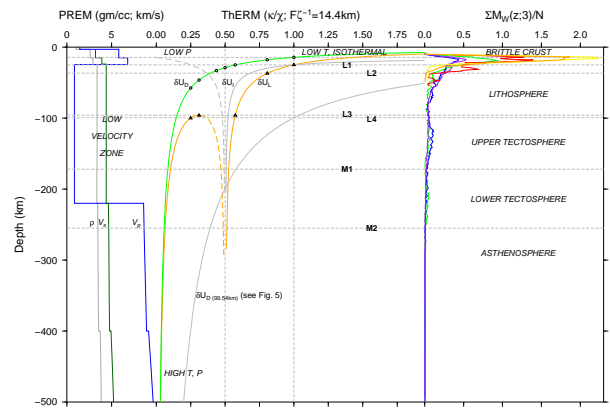}
\caption{Upper mantle correlation of: (left) Earth's radial elastic structure \protect{\cite{Dziewonski1981}}; (center) ThERM, scaled to $F\zeta^{-1}=14.4 km$ \protect{\cite{Patton2009,Patton2010}}, and (right) event normalized depth-moment release curves $\Sigma{M}_{W}\slash{N}$ for convergent ocean-ocean (purple), convergent ocean-continent (blue), convergent continent-continent (green), divergent continent-continent (red), divergent ocean-ocean (orange), and transcurrent (yellow) margins.}
\end{figure}

\section{Discussion}
\subsection{General hypotheses}
The depth to which differential stresses can persist in any terrestrial planet depends on temperature, pressure, and time, through complicated solid-state interactions of matter and energy.  However, given that terrestrial planets are spheroidally-shaped, reflecting the combined effects of their self-gravity and rotational momentum, the magnitude of these differential stresses must be small.  Furthermore, given that temperature is expected to increase with depth (Figure 2b), and that microphysical mechanisms for solid-state creep are thermally activated, even these small differential stresses must diminish rapidly with depth.  In the static case, they can be expected to decay entirely, leaving only a radially decreasing static pressure to hold off collapse.  Consequently, the maintenance of differential stresses at depth in a planet requires some dynamic process \cite{McKenzie1967}.  Hence, it is natural to think of terrestrial mantles as heat engines, the evolution of which is governed by a competition between convection and diffusion processes.  For more than 40 years, geodynamic models have conformed to the assumptions of the standard viscous Earth model \cite{Bercovici2000}.  However, it is likely that these processes are more interesting than heretofore recognized.

With pressure and temperature as boundary conditions on a self-gravitating planet, and confining pressures much larger than potential differential stresses, it is hard to argue that material strength matters, except for the dynamic rescaling theorem \cite{Patton2010}, which predicts that shear localization focuses on the smallest crystalline structures of a solid system.  This is necessary, so that the global dissipation of energy is minimized.  This same logic applies equally to the dynamic rupture models of strong-motion seismology \cite{Barenblatt1959,Ida1972}, and the behavior of Earth's plate-like toroidal motions \cite{Bercovici1995}.

An immediate consequence of this prediction is that shear waves can be propagated throughout a thermomechanical mantle, whereas in a viscous one, no such propagation is possible, except perhaps at ultrasonic speeds.  Absent this theorem, however, one must accept the geodynamicist's approximation, that over long time and length scales the mantle is effectively viscous.  As plausible as this might sound, it is impossible to falsify.  Moreover, it is inconsistent with the fact that rocks loaded in the laboratory exist in the solid-state.  Thus, it is clear that the failure of the standard Earth model arises solely from a theoretical deficiency.  On the other hand, with the dynamic rescaling theorem, the only substantive differences between deformation of a rock sample in the laboratory, and tectonic deformation of the Earth, are the relative magnitude of the confining pressure and effect of global conservation laws.  In other words, shear localization at the planetary-scale must account for the incompatibility of rectilinear motions with the spheroidally curved geometry of the planet itself, while in the laboratory this is of no concern. 

The thermodynamics of non-linear elastic DG-2 materials is spatiotemporally invariant.  Therefore, a thermomechanical Earth model can be formed simply by scaling up to Earth's radial structure and applying pressure and temperature boundary conditions at its surface \cite{Patton2009,Patton2010}.  The result immediately predicts a variation of pressure and temperature expected for terrestrial mantles.  Furthermore, it predicts that the outer colder parts of the mantle should be thermomechanically rigid, thermodynamically isothermal, and subject to brittle shear localization, while the deeper hotter parts should be thermomechanically ductile and thermodynamically adiabatic.  Adiabaticity prevails as $\smash{\frac{\kappa}{\chi}\longrightarrow{0}}$, consistent with depths in the lower mantle, and coincidentally where Birch \cite{Birch1952} showed it to pertain on the basis of observed seismic wave speed variations.  The asthenosphere of a thermomechanical Earth is not adiabatic, because differential normal stresses there are large enough to cause shear localization.  Moreover, the vanishing of seismicity at the asthenosphere-mesosphere boundary, at about $690 km$ depth (Fig. 5), reflects this fundamental change in thermodynamic conditions.  For comparison, the largest variations in earthquake depth-moment release (Fig. 6) occur in the low velocity zone, at depths consistent with Earth's lithosphere and tectosphere.

For a self-gravitating solid body, like a terrestrial planet, we can anticipate some degree of interplay between the inhomogeneous statistical distribution of length scales in the body, and the distribution of thermal lengths over which the pressure gradients might act.  This interplay is expressed particularly in the structure of the thermomechanical boundary layer that forms adjacent to the cold surface of the planet, but also by the fact that elastic shear waves are propagated throughout the mantle.  Thus, in order for any portion of a thermomechanical planet to suffer deformation, there must be measurable contrasts in material competence.  Furthermore, the dislocation diffusivity $\chi$ must be greater than the thermal diffusivity $\kappa$ in deforming portions of this complex system.  Consequently, the thermomechanical boundary layer that forms will always be $\zeta$ times thicker than the purely thermal one.  Depth and regional variations in seismic moment release for such a planet are therefore to be expected.  Finally, given that pressure and temperature variations are explicitly predicted by theory, it is reasonable to suppose that variations in material competence will exhibit strong dependencies on bulk composition and volatile content.     

\subsection{Seismicity}
\subsubsection{Source dynamics}
The foregoing discussion suggests that the abrupt cutoff of seismicity at about $690 km$ depth in the convecting mantle is the result of an entropic transition from spontaneous supersonic to forced subsonic rupture, with increasing pressure and temperature (Fig. 5).  If so, some component of supershear must be present in all seismically observed ruptures.
Furthermore, this implies that mass transport between the upper and lower mantle can be consistent with both plate tectonics and the global distribution of earthquakes.  This hypothesis differs from that of \cite{Isacks1968}, in which descending slabs containing seismic sources are barred, mechanically, from penetrating into the lower mantle due to an abrupt increase in viscosity.

\subsubsection{Attenuation in the upper mantle}
Based on the work of Dunn \& Rajagopal \cite{Dunn1995}, normal stress effects proportional to $\beta_1$ and $\beta_2$ are expected to have the same order of magnitude.  As shown above, the effects of thermomechanical competence are manifest throughout the mantle, but play a central role in the observed structure of Earth's thermomechanical boundary layer.  Consequently, we can expect that non-linear viscous effects, proportional to $\beta_2$, might also be expressed in the boundary layer.  Coincidentally, shear Q is low in upper mantle, between $80-670 km$ depths \cite{Dziewonski1981}, and exhibits large lateral variations in the crust and uppermost mantle correlated with past and present tectonic activity \cite{Romanowicz2000}.

\subsection{Gravity and topography}
\subsubsection{Strength heterogeneity}
Thermomechanical competence is a non-dimensional measure of material strength.  Consequently, at low temperatures near the surface of a terrestrial planet, strength heterogeneities can be 'frozen' into the crust and mantle, which then support topography.  The depth cutoff for this strength heterogeneity is approximated by mode $H4$ of ThERM for continental regions and mode $L4$ for oceanic regions.  Variations in thermomechanical competence, however, can persist to even greater depths, and possibly as deep as mode $M4$.  Corresponding depths for Earth are 14.4 km, 100 km, and 690 km, respectively.  Coincidentally, this range of depths is similar to the observed size range of SSSBs in the inner solar system.

\subsubsection{Density heterogeneity}
Sedimentation of entrained particles, either from a gas or liquid is an important source of density heterogeneities near the surface of terrestrial planets.  So too is magmatic differentiation.  As shown above, however, it is possible that source regions for common magma types are local subsystems, confined within the body of the planet, in which classical thermodynamic conditions prevail.  If so, tectonic deformation driven by coupled poloidal and toroidal motions can be expected to produce density heterogeneities throughout the crust and upper mantle.   

\subsubsection{Hypothetical wavebands}
Wavebands for the interpretation of gravity and topography observations on terrestrial planets (e.g. \cite{Wieczorek2007}) can be estimated using the Cartesian wavelengths for fold-like perturbations in the surface layers of ThERM.  For each band, the minimum and maximum wavelengths are given by the $2\times$ and $\zeta\times$ multiples of the depths to the isobaric shears, consistent with the material wave solutions admitted by the diharmonic equation (Fig. 4a, solid blue curve).  Translating these wavelengths $\lambda$ to spherical harmonic degree $l$ using Jean's relation $\smash{\lambda=\frac{2\pi{r}}{l+\frac{1}{2}}}$, where $r$ is planetary radius, these bands can be compared directly to admittance and correlation spectra.  Presently, only Earth's radial structure is known well enough to allow this comparison (Fig. 1).  The correlation shows promise, but detailed consideration is left for another day.

\section{Conclusion}
Patton and Watkinson \cite{Patton2013} recently reported that equations \eqref{diharmonic} and \eqref{rescaling_non-dim} were powerful tools for understanding plate tectonics and structural geology on Earth.  This paper further supports that claim by demonstrating that diharmonic rescaling is a non-linear attractor for solid-state deformation in the toroidal motions of terrestrial planets generally.  Testing of these hypothesis should proceed forthwith.

\bibliography{References_PATTON}{}
\bibliographystyle{model6-num-names}
\biboptions{sort&compress}
\end{document}